\documentclass[conference, 10pt]{IEEEtran}

\pdfoutput=1

\usepackage{xspace}
\usepackage{todonotes}

\usepackage{tikz}
\usepackage{pgfplots}
\usepackage{tikzscale}
\pgfplotsset{compat=newest}
\usetikzlibrary{plotmarks}
\usepackage{rotating}
\usepackage[absolute,overlay]{textpos}
\usepackage{circuitikz}

\usepackage{amsmath}
\usepackage{amssymb}
\usepackage{amsthm}

\newcommand{\R}{\mathbb{R}}

\newcommand{\fpart}[2]{\frac{\partial #1}{\partial #2}}

\usepackage{hyperref}

\usepackage{hf-tikz}
\usetikzlibrary{ 
	calc,
	arrows,
	arrows.meta,
	automata, 
	shapes, 
	snakes, 
	positioning, 
	decorations,
	decorations.text,
	fit,
	matrix,
	mindmap
}

\usetikzlibrary{calc}
\usetikzlibrary{decorations.pathreplacing,decorations.markings,shapes.geometric}
\usetikzlibrary{shapes}
\usetikzlibrary{decorations.pathreplacing}

\tikzset{radiation/.style={{decorate,decoration={expanding waves,
angle=90,segment length=4pt}}}}
\tikzset{relay/.pic={
            code={\tikzset{scale=5/10}
            \draw[semithick] (0,0) -- (1,4);
            \draw[semithick] (3,0) -- (2,4);
            \draw[semithick] (0,0) arc (180:0:1.5 and -0.5)
            node[above, midway]{#1};
            \node[inner sep=4pt] (circ) at (1.5,5.5) {};
            \draw[semithick] (1.5,5.5) circle(8pt);
            \draw[semithick] (1.5,5.5cm-8pt) -- (1.5,4);
            \draw[semithick] (1.5,4) ellipse (0.5 and 0.166);
            \draw[semithick,radiation,decoration={angle=45}]
            (1.5cm+8pt,5.5) -- +(0:2);
            \draw[semithick,radiation,decoration={angle=45}] 
            (1.5cm-8pt,5.5) -- +(180:2);}}
        }

\makeatletter
\tikzset{
    database/.style={
        path picture={
            \draw (0, 1.5*\database@segmentheight) circle [x radius=\database@radius,
            y radius=\database@aspectratio*\database@radius];
            \draw (-\database@radius, 0.5*\database@segmentheight)
            arc [start angle=180,end angle=360,x radius=\database@radius,
            y radius=\database@aspectratio*\database@radius];
            \draw (-\database@radius,-0.5*\database@segmentheight)
            arc [start angle=180,end angle=360,x radius=\database@radius,
            y radius=\database@aspectratio*\database@radius];
            \draw (-\database@radius,1.5*\database@segmentheight)
            -- ++(0,-3*\database@segmentheight) arc [start angle=180,end angle=360,
            x radius=\database@radius, y radius=\database@aspectratio*\database@radius]
            -- ++(0,3*\database@segmentheight);
        },
        minimum width=2*\database@radius + \pgflinewidth,
        minimum height=3*\database@segmentheight + 2*\database@aspectratio*\database@radius + \pgflinewidth,
    },
    database segment height/.store in=\database@segmentheight,
    database radius/.store in=\database@radius,
    database aspect ratio/.store in=\database@aspectratio,
    database segment height=0.1cm,
    database radius=0.25cm,
    database aspect ratio=0.35,
}
\makeatother

\usepackage{graphicx}
\usepackage{amsmath}
\usepackage{array}
\usepackage[caption=false,font=footnotesize]{subfig}
\usepackage{microtype}
\usepackage{dblfloatfix}
\usepackage{cite}
\usepackage{algorithm, algorithmic}
\usepackage{ctable}
\usepackage{siunitx}

\begin{document}
\title{Energy-Efficient Edge-Facilitated Wireless Collaborative Computing
using Map-Reduce}

\author{%
\IEEEauthorblockN{%
    Antoine Paris,
	Hamed Mirghasemi,
	Ivan Stupia and
	Luc Vandendorpe}
\IEEEauthorblockA{%
	ICTEAM/ELEN/CoSy, UCLouvain, Louvain-la-Neuve, Belgium\\
	Email: \{%
	\href{mailto:antoine.paris@uclouvain.be}{antoine.paris},
	\href{mailto:seyed.mirghasemi@uclouvain.be}{seyed.mirghasemi},
	\href{mailto:ivan.stupia@uclouvain.be}{ivan.stupia},
	\href{mailto:luc.vandendorpe@uclouvain.be}{luc.vandendorpe}\}%
    @uclouvain.be
}}

\maketitle

\begin{abstract}
    In this work, a heterogeneous set of wireless devices sharing a
    common access point collaborates to perform a set of tasks.
    Using the Map-Reduce distributed computing framework, the tasks
    are optimally distributed amongst the nodes with the objective of
    minimizing the total energy consumption of the nodes while satisfying
    a latency constraint.
    The derived optimal collaborative-computing scheme takes
    into account both the computing capabilities of the nodes and the
    strength of their communication links.
    Numerical simulations illustrate the benefits of the proposed
    optimal collaborative-computing scheme over a blind
    collaborative-computing scheme and the non-collaborative scenario,
    both in term of energy savings and achievable latency.
    The proposed optimal scheme also exhibits the interesting feature
    of allowing to trade energy for latency, and vice versa. 
\end{abstract}

\begin{IEEEkeywords}
    wireless collaborative computing, distributed computing,
    Map-Reduce, energy-efficiency, fog computing.
\end{IEEEkeywords}

\section{Introduction}
We consider a set of $K$ nodes, indexed by the letter%
~\mbox{$k \in [K]$}, sharing a common access
point (AP), base station (BS) or gateway in the context of
low-power wide-area networks (LPWAN).
A node can be any device able to wirelessly communicate with the AP
and perform local computations.
Under a given latency constraint~$\tau$, each node $k$ wants to compute
a certain function $\phi(d_k, w)$ where~$d_k \in [0, 1]^D$ is
some $D$-bit local information available to node $k$ (e.g., sensed
information or local state) and $w \in [0, 1]^L$ is a $L$-bit file
with $L \gg D$ bits (e.g., a dataset) that might, for instance,
be cached at the AP~\cite{caching}.
In the context of smart cities or smart buildings, $w$ could be the
result of the aggregation over space and time of information
sensed from the environment through a network of sensors (e.g.,
traffic density or temperature) whereas the nodes could be actuators
having some local state $d_k$ that periodically need to perform some
latency-sensitive computations to decide whether to take some actions (e.g.,
smart traffic lights or smart thermostats). Other applications
include fog computing, mobile crowd-sensing or wireless distributed
systems.

Owing to the unacceptable delay of mobile cloud computing (MCC), and
in the absence of a mobile edge computing (MEC) server nearby,
the computing and storage capabilities of wireless
devices are limited. It might thus be the case, for example,
that $w$ is too large to fit in the memory of a single node, or that
the nodes are not individually powerful enough to satisfy the
latency constraint. To overcome those limitations, a
collaborative-computing scheme based on the Map-Reduce distributed
computing framework~\cite{MapReduce} is proposed. This distributed
computing model involves local computations at the nodes and
communication between the nodes
via the~AP (i.e., the edge of the network is \textit{facilitating}
the communication between the nodes). In some applications, one
could also deliberately avoid the use of a third-party owned MCC
or MEC for privacy reasons.

The problem setup and distributed computing model used in this work
essentially follows~\cite{scalableWDC}, with the exceptions that
we consider the set of nodes to be \textit{heterogeneous} in term 
of computing capabilities and channel strengths and add an explicit
latency constraint.
%and that we allow $w$
%to be arbitrarily divided and distributed to the nodes.
Prior works on wireless distributed computing using Map-Reduce,
e.g.,~\cite{scalableWDC, coding-edge, comm-comp-tradeoff, wireless-mr},
mainly focus on \textit{coded distributed computing} (CDC) and study
the trade-off between the computation and communication loads
incurred by the collaboration. 
Motivated by the fact that wireless devices are often limited in
energy and that most computing tasks are accompanied by a
latency constraint, this work shifts focus towards optimizing
the collaborative-computing scheme to minimize the total energy
consumption of the nodes, while satisfying the latency constraint.
To our knowledge, this work is the first to
incorporate those considerations in a Map-Reduce based wireless
collaborative-computing scheme.

Throughout this paper, we assume that there is some central entity
having perfect knowledge of the channel state information (CSI)
and computing capabilities of all the nodes that coordinates
the collaboration.

Section~\ref{sec:system-model} starts by describing in details
the distributed computing model and the energy and time consumption
models for both local computation and communication between the
nodes. Next, Sec.~\ref{sec:problem-formulation} formulates the
problem as an optimization problem that turns out to be convex
and to have a semi-closed form solution, given in
Sec.~\ref{sec:optimal-sol}. 
Section~\ref{sec:num-exp} then benchmarks the performances of
the optimal collaborative-computing scheme against a blind
collaborative-computing scheme and the non-collaborative scenario
through numerical experiments.
Finally, Sec.~\ref{sec:ccl} discusses the results obtained in this
work and opportunities for future research.

%\begin{figure}
%    \centering
%    \input{figures/problem-setup.tex}
%    \caption{Illustration of the problem setup.}
%	\label{fig:problem-setup}
%\end{figure}

\section{System model}
\label{sec:system-model}
This section details the distributed computing model used
in this work, namely Map-Reduce, and quantifies the time
and energy consumed by each phase of the collaboration.

\subsection{Distributed computing model}
\label{sec:distributed-comp-model}
The tasks are shared between the $K$ nodes according to
the Map-Reduce framework~\cite{MapReduce}. First, we assume that
the file $w$ can be arbitrarily divided in $K$ smaller files
$w_k$ (one for each node) of size $l_k \in \R_{\ge 0}$ bits%
\footnote{In practice, $l_k$ should be an integer multiple
of the size of the smallest possible division of $w$.
In this work, we relax this practical consideration to
avoid dealing with integer programming later on. Note that $l_k = 0$
is also possible, in which case node $k$ does not participate to
the collaboration.}
such that $w_k \cap w_l = \emptyset$ for all $k \neq l$
and $w = \bigcup_{k=1}^K w_k$. We neglect the time
and energy needed to transmit $w_k$
from the AP to node $k$, for all $k \in [K]$.

To make collaboration between the nodes possible, we also
assume that the local data $\{d_k\}_{k=1}^K$ were shared
between all the nodes through the AP in a prior phase
that we neglect in this work because $D$ is assumed to be
relatively small.

During the first phase of the Map-Reduce framework, namely
the \textit{Map phase}, each node $k$ computes intermediate values
\[ \textstyle{v_{k,l} = g_k(d_l, w_k), \qquad l \in [K]} \]
where $g_k: [0,1]^D \times [0, 1]^{l_k} \to
[0, 1]^{(l_k/L)T}$ is the \textit{Map function}
executed at node $k$. The size (in bits) of the intermediate
values produced at node $k$ is assumed to be proportional to
$l_k$. Each node $k$ thus computes intermediate values for
all the other nodes (i.e., $v_{k,l}$ for all $l \neq k$)
and for itself (i.e., $v_{k,k}$) using the part $w_k$
of $w$ received from the AP.

Next, the nodes exchange intermediate values with each
other in the so-called \textit{Shuffle phase}.
More precisely, each node $k$ transmits the intermediate
values $v_{k,l} = g_k(d_l, w_k)$ to node $l$ via the AP,
for all $l \neq k$. In total, node $k$ thus needs to transmit
$(K-1)(l_k/L)T$ bits of intermediate values to the AP.

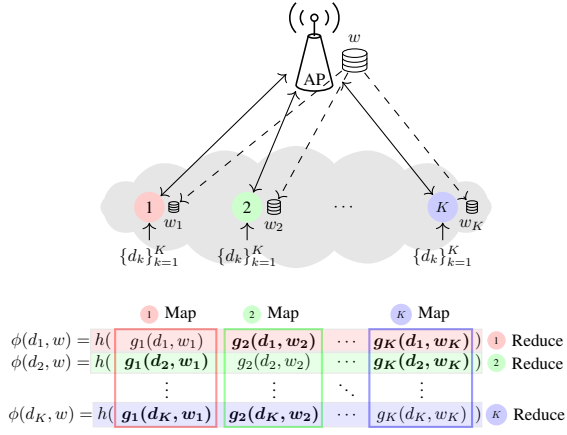
\begin{figure}
    \centering
    % FIXME: try to put back the scale at 0.75 at the end if space allows it
\begin{tikzpicture}[scale=0.65, every node/.style={scale=0.65}]
    \node[database, label=above:$w$] at (1.2, 0) (w) {};
    \node[cloud, fill=gray!20, cloud puffs=16, cloud puff arc=100,
    minimum width=8cm, minimum height=2.5cm, aspect=1] at (0, -3) {};
    
    \pic[draw, scale=0.5, local bounding box=AP] at (0, -0.5)
    {relay=AP};

    \node[circle, fill=red!20] at (-3, -3) (1) {1};
    \node[database, database radius=0.1cm,
        database segment height=0.05cm,
        label=below:$w_1$] at (-2.5, -3) (w1) {};
    \node[circle, fill=green!20] at (-1, -3) (2) {2};
    \node[database, database radius=0.13cm,
        database segment height=0.07cm,
        label=below:$w_2$] at (-0.45, -3) (w2) {};
    \node at (1, -3) () {$\cdots$};
    \node[circle, fill=blue!20, scale=0.85] at (3, -3) (K) {$K$};
    \node[database, database radius=0.11cm,
        database segment height=0.06cm,
        label=below:$w_K$] at (3.6, -3) (wK) {};

    \draw[<->] (1) -- (AP);
    \draw[<->] (2) -- (AP);
    \draw[<->] (K) -- (AP);

    \node at (-3, -4) (d1) {$\{d_k\}_{k=1}^K$};
    \draw[->] (d1) -- (1);
    \node at (-1, -4) (d2) {$\{d_k\}_{k=1}^K$};
    \draw[->] (d2) -- (2);
    \node at (3, -4) (dK) {$\{d_k\}_{k=1}^K$};
    \draw[->] (dK) -- (K);

    \draw[->, dashed] (w) -- (w1);
    \draw[->, dashed] (w) -- (w2);
    \draw[->, dashed] (w) -- (wK);
    \node at (0, -6.5) {%
            $\begin{matrix}
                g_1(d_1, w_1)   & \boldsymbol{g_2(d_1, w_2)} & \cdots &
                \boldsymbol{g_K(d_1, w_K)} \\
                \boldsymbol{g_1(d_2, w_1)}   & g_2(d_2, w_2) & \cdots &
                \boldsymbol{g_K(d_2, w_K)} \\
                \vdots          & \vdots        & \ddots & \vdots \\
                \boldsymbol{g_1(d_K, w_1)}   &
                \boldsymbol{g_2(d_K, w_2)} & \cdots & g_K(d_K, w_K)
            \end{matrix}$
    };

    \draw[color=red!40, thick] (-3.7, -5.5) rectangle (-1.63, -7.5);
    \node[circle, fill=red!20, label=right:Map, scale=0.6] at
        (-3, -5.2) () {1};   
    \draw[color=green!40, thick]  (-1.46, -5.5) rectangle (0.53, -7.5);
    \node[circle, fill=green!20, label=right:Map, scale=0.6] at
        (-0.9, -5.2) () {2};   
    \draw[color=blue!40, thick]   (1.5, -5.5) rectangle (3.6, -7.5);
    \node[circle, fill=blue!20, label=right:Map, scale=0.6] at
        (2.15, -5.2) () {$K$};   
    
    \draw[fill=red, opacity=0.1] (-4.2, -5.5) rectangle (3.8, -5.95);
    \draw[fill=green, opacity=0.1] (-4.2, -5.98) rectangle (3.8, -6.4);
    \draw[fill=blue, opacity=0.1] (-4.15, -7) rectangle (3.8, -7.5);

    \node at (-4.8, -5.73) {$\phi(d_1, w) = h($};
    \node at (3.73, -5.73) {$)$};
    \node[circle, fill=red!20, scale=0.6, label=right:Reduce]
        at (4.1, -5.73) () {1};   
    \node at (-4.8, -6.19) {$\phi(d_2, w) = h($};
    \node at (3.73, -6.19) {$)$};
    \node[circle, fill=green!20, scale=0.6, label=right:Reduce]
        at (4.1, -6.19) () {2};  
    \node at (-4.8, -7.25) {$\phi(d_K, w) = h($};
    \node at (3.73, -7.25) {$)$};
    \node[circle, fill=blue!20, scale=0.6, label=right:Reduce]
        at (4.1, -7.25) () {$K$};   
\end{tikzpicture}
    \caption{Illustration of the the Map-Reduce distributed computing model.
    The computation of $\{\phi(d_k, w)\}_{k=1}^K$ is distributed between $K$
    nodes. During the Map phase, each node $k$ computes intermediate values
    $\{g_k(d_l, w_k)\}_{l=1}^K$ (see framed columns on the above figure).
    Next, during the Shuffle phase, the intermediate values in bold on the
    figure are transmitted via the AP to the nodes for which they have been
    computed. Finally, during the Reduce phase, each node $l$ combines the
    intermediate values $\{g_k(d_l, w_k)\}_{k=1}^K$ to obtain
    $\phi(d_l, w)$ (see colored rows on the above figure).}
    \label{fig:map-reduce}
\end{figure}

Finally, during the \textit{Reduce phase}, each node $l$
combines the $T$ bits of intermediate values
$\{v_{k,l} = g_k(d_l, w_k)\}_{k=1}^K$ as
\[ \phi(d_l, w) = h(g_1(d_l, w_1), g_2(d_l, w_2), \dots, g_K(d_l, w_K)) \]
where $h: [0,1]^T \to [0, 1]^O$ is the \textit{Reduce
function}. The Map-Reduce distributed computing model is illustrated
in Fig.~\ref{fig:map-reduce}.

\subsection{Local computing model}
\label{sec:local-comp-model}
During the Map and the Reduce phases, the nodes have to perform some
local computations. The local computing model used in this work
follows~\cite{EERAMECO}.
The number of CPU cycles required to process 1-bit of
input data is noted $C_k$ while the amount of energy consumed per CPU cycle
is noted $P_k$. The amounts of energy consumed at node $k$ during the Map
and the Reduce phases\footnote{Note that we assume that $C_k$ and $P_k$
are the same for both phases.} are thus given by
\begin{equation}
    E^\text{MAP}_k = (KD + l_k)C_kP_k
    \quad \text{and} \quad
    E^\text{RED}_k = TC_kP_k,
    \label{eq:e-map-red}
\end{equation}
respectively.
Next, letting $F_k$ be the number of CPU cycles per second at node $k$, the
amounts of time required for the Map and the Reduce phases are given by
\begin{equation}
    t^\text{MAP}_k = (KD + l_k)C_k/F_k
    \quad \text{and} \quad
    t^\text{RED}_k = TC_k/F_k,
    \label{eq:t-map-red}
\end{equation}
respectively.
One can already observe that we can control the energy and time consumed
at node $k$ by the Map phase through the variable $l_k$.
At the opposite, we don't have any control on the energy and time
consumed by the Reduce phase. As a consequence, and because the Map and the
Shuffle phases must be over before the Reduce phase can start, the time
remaining for the Map and the Shuffle phases is given by
\mbox{$\tau - \max_k\{t^\text{RED}_k\}$}, i.e., the slowest node reduces
the available time $\tau$ by the amount of time it needs for the Reduce
phase.

\subsection{Communications from the nodes to the AP}
\label{sec:comm-model}
During the Shuffle phase, nodes exchange intermediate values through
the AP\@. This exchange thus involves both an uplink communication (nodes
to AP) and a downlink communication (AP to nodes).
In most applications however, it is reasonable to assume that the downlink
rates are much larger than the uplink rates. For this reason, we neglect the
time needed for the downlink communication in this work. 

We assume that all the nodes can communicate in an orthogonal manner to
the AP (e.g., through frequency division multiple access techniques).
We also make the common assumption that the allowed latency
$\tau$ is smaller than the channel coherence time.
Let $h_k \in \mathbb{C}$ denote the wireless channel from node $k$ to
the AP, $p_k$ the RF transmit power of node $k$,
$B$ the communication bandwidth,
$\sigma^2$ the noise power at the AP in the bandwidth $B$, and
$\Gamma$ the SNR gap.
The achievable uplink rate of node $k$ is then given by
\begin{equation*}
    \textstyle{%
        r_k(p_k) = B\log_2\left(1 + \frac{p_k|h_k|^2}{\Gamma \sigma^2}
        \right).
    }
\end{equation*}
The time required by node $k$ to transmit the $(K-1)(l_k/L)T$ bits
of intermediate values to the AP is thus given by
\mbox{$t^{\text{SHU}}_k = \alpha l_k/r_k(p_k)$}
where $\alpha = (K-1)T/L$ has been defined to ease notations.
Then, inspired by~\cite{EERAMECO}, we define
$f(x) = \sigma^2\Gamma(2^{x/B} - 1)$, and write the energy consumed at node
$k$ to transmit the intermediate values as
\begin{equation}
    \textstyle{%
    E^{\text{SHU}}_k = p_kt^\text{SHU}_k =
    \frac{t^\text{SHU}_k}{|h_k|^2}f\left(
    \frac{\alpha l_k}{t^\text{SHU}_k}\right).
    }
    \label{eq:Eshu}
\end{equation}
Through the variables $l_k$ and $p_k$
(or, equivalently, $t^\text{SHU}_k$),
we thus have control on the energy and time consumed at node $k$ during
the Shuffle phase. 
%This rather intricate way of defining $E^\text{SHU}_k$ as a function of
%$t^\text{SHU}_{k}$ only is inspired by

\section{Problem formulation}
\label{sec:problem-formulation}
As mentioned in the introduction, the objective is to optimize the
collaborative-computing scheme to minimize the total energy
consumption of the nodes, while satisfying the latency constraint $\tau$.
This can be mathematically formulated as follows
\begin{align}
    & \underset{\{l_k\}, \{t^\text{SHU}_k\}}{\text{minimize}}
    && \textstyle{\sum_{k=1}^K E^\text{MAP}_k + E^\text{SHU}_k
    + E^\text{RED}_k} &
    \nonumber\\
    & \text{subject to} 
    && l_k, t^\text{SHU}_k \geq 0,\quad k \in [K]
    \nonumber\\
    &&& t^\text{MAP}_k + t^\text{SHU}_k \le \tau -
    \textstyle{\max_k\{t^\text{RED}_k\}}, \quad k \in [K] 
    \label{const:lat-1} \\
    &&& \textstyle{\sum_{k=1}^K l_k = L}. &
    \label{const:compl-1}
\end{align}
Constraint~\eqref{const:lat-1} directly follows from the discussion
at the end of Sec.~\ref{sec:local-comp-model} while%
~\eqref{const:compl-1} ensures that the partition $\{w_k\}_{k=1}^K$
of $w$ fully covers $w$.
Substituting Eqs.~\eqref{eq:e-map-red}-\eqref{eq:Eshu} in the above
optimization problem and removing the constant terms from the objective
function, we obtain
\begin{align}
    & \underset{\{l_k\}, \{t^\text{SHU}_k\}}{\text{minimize}}
    && \textstyle{\sum_{k=1}^K l_kC_kP_k+ \frac{t^\text{SHU}_k}{|h_k|^2}
    f\left(\frac{\alpha l_k}{t^\text{SHU}_k}\right)} &
    \label{eq:obj-fct-1} \\
    & \text{subject to} 
    && l_k, t^\text{SHU}_k \geq 0,\quad k \in [K] 
    \nonumber \\
    &&& l_k\frac{C_k}{F_k}
    + t^\text{SHU}_k \le \tau_k, \quad k \in [K]
    \label{const:lat-2} \\
    &&& \textstyle{\sum_{k=1}^K l_k = L} \nonumber &
\end{align}
with $\tau_k$, the \textit{effective latency constraint}
of node $k$, given by
\begin{equation}
    \textstyle{%
    \tau_k = \tau - T\max_k\{C_k/F_k\} - KDC_k/F_k.}
    \label{eq:tauk}
\end{equation}
This last optimization problem is very
similar to the one formulated in~\cite{EERAMECO} and is known to be
convex~\cite[Lemma 1]{EERAMECO}.

Next, one can observe that the objective function~\eqref{eq:obj-fct-1}
is always decreasing with $t^\text{SHU}_k$. Indeed, for a fixed number
of bits $\alpha l_k$ to transmit during the Shuffle phase,
increasing the duration of the transmission
$t^\text{SHU}_k$ always decreases the energy consumption $E^\text{SHU}_k$.
As a consequence, constraint~\eqref{const:lat-2} is always active at the
optimum and can thus be turned into an equality
constraint\footnote{In the particular case where $l_k = 0$, the value of
$t^\text{SHU}_k$ does not impact the objective function and imposing
$t^\text{SHU}_k = \tau_k$ to make the constraint active is thus not an
issue.}.
We can thus get rid of half of the optimization variables
by substituting $l_k$ by
$\frac{F_k}{C_k}(\tau_k - t^\text{SHU}_k)$. This leads to
\begin{align}
    & \underset{\{t^\text{SHU}_k\}}{\text{minimize}}
    && \textstyle{\sum_{k=1}^K (\tau_k - t^\text{SHU}_k)F_kP_k} &
    \nonumber \\
    &&& \textstyle{+ \frac{t^\text{SHU}_k}{|h_k|^2}
        f\left(\alpha\frac{F_k}{C_k}
            \left(\frac{\tau_k}{t^\text{SHU}_k} - 1\right)
        \right)}
    \nonumber \\
    & \text{subject to} 
    && 0 \leq t^\text{SHU}_k \leq \tau_k, \quad k \in [K]
    \nonumber \\
    &&& \textstyle{\sum_{k=1}^K \frac{F_k}{C_k}(\tau_k
    - t^\text{SHU}_k) = L}. &
    \label{const:lat-3}
\end{align}

\section{Optimal solution}
\label{sec:optimal-sol}
We start by defining the partial Lagrangian as follows
\[ 
    \begin{split}
        \textstyle{\mathcal{L}(\{t_k\}, \lambda) =
        \sum_{k=1}^K (\tau_k - t_k)F_kP_k 
        + \frac{t_k}{|h_k|^2}f\left(
            \alpha\frac{F_k}{C_k}\left(\frac{\tau_k}{t_k}
            - 1\right)\right)} \\
    \textstyle{+ \lambda\left(L - \sum_{k=1}^K \frac{F_k}{C_k}
    (\tau_k - t_k)\right)}
    \end{split}
\]
where $t^\text{SHU}_k$ has been replaced by $t_k$ to ease notations
and with $\lambda$ the Lagrange multiplier associated
to~\eqref{const:lat-3}.
Then, applying the KKT conditions to the partial Lagrangian leads to
\begin{align*}
    \textstyle{\fpart{\mathcal{L}}{t_k}\big|_*} &=
    \textstyle{-F_kP_k +
    \frac{1}{|h_k|^2}f\left(\alpha\frac{F_k}{C_k}\left(
    \frac{\tau_k}{t^*_k} - 1\right)\right)} \\
    & -
    \textstyle{\frac{\alpha}{|h_k|^2}\frac{F_k}{C_k}
    \frac{\tau_k}{t^*_k}
    f'\left(\alpha\frac{F_k}{C_k}\left(\frac{\tau_k}{t^*_k}
        - 1\right)\right)
    + \lambda^*\frac{F_k}{C_k}} \\
    &= \textstyle{-F_kP_k - \frac{\Gamma \sigma^2}{|h_k|^2}
    + \lambda^*\frac{F_k}{C_k}} \\
    & \textstyle{+ \frac{\Gamma \sigma^2}{|h_k|^2}\left(1 - \alpha
    \frac{\ln(2)}{B}
	\frac{F_k}{C_k}\frac{\tau_k}{t^*_k}\right)
	2^{\frac{\alpha}{B}\frac{F_k}{C_k}\left(\frac{\tau_k}{t^*_k}
    - 1\right)}}\\
	&
    \textstyle{%
	\begin{cases}
		> 0, & t^*_k = 0 \\
        = 0, & t^*_k \in]0, \tau_k] \\
		< 0, & t^*_k = \tau_k \Rightarrow l^*_k = 0,
\end{cases}}
\end{align*}
with
\[
    \textstyle{%
    \sum_{k=1}^K \frac{F_k}{C_k}(\tau_k - t^*_k) = L}.
\]
The first case (i.e., $> 0$) can't happen as the objective goes to
$+\infty$ when $t_k$ goes to 0.
The last case (i.e., $< 0$) tells us when a node
does not participate in the Map and Shuffle phases (i.e., when $l^*_k = 0$).
It can be re-written as
\begin{equation}
    \textstyle{%
    C_kP_k + \alpha\frac{\Gamma \sigma^2}{|h_k|^2}\frac{\ln(2)}{B}
    > \lambda^*.}
    \label{eq:threshold}
\end{equation}
The left-hand side of the inequality corresponds to the marginal energy
consumption of node $k$ per bit received, when node $k$ hasn't received
any bit yet, i.e., at $l_k = 0$. Indeed, the first term corresponds to
the marginal energy consumption incurred by the Map phase while the second
term corresponds to the marginal energy consumption incurred by the Shuffle
phase. In other words, the left-hand side of~\eqref{eq:threshold}
can be interpreted as the ``price to start collaborating''.
If this price is greater than a threshold given by $\lambda^*$, then
$l^*_k = 0$, meaning that node $k$ does not participate to the Map and
Shuffle phases\footnote{Note that it still participates to the Reduce
phase as it still needs to obtain $\phi(d_k, w)$.}.
%This gives a threshold-based policy to decide
%which nodes effectively participate to the collaboration.
Finally, solving the remaining case (i.e., $= 0$) for $t^*_k$ leads to
\begin{equation}
    \textstyle{%
	t^*_k =
	\frac{%
        \alpha\frac{\ln(2)}{B}\frac{F_k}{C_k}
	    \times\tau_k
	}
	{%
		W_0\left\{\frac{1}{e}
		\left(\frac{|h_k|^2}{\Gamma \sigma^2}\frac{F_k}{C_k} 
			\left(\lambda^*- C_kP_k\right)
	- 1\right)e^{\alpha\frac{\ln(2)}{B}\frac{F_k}{C_k}}\right\}
	+ 1
	}
    }
    \label{eq:opt}
\end{equation}
where $W_0(\cdot)$ is the main branch of the Lambert function.
% FIXME: missing in \ge \lambda^* above
%Note: plugging $\lambda^* = KC_kP_k +
%\alpha\frac{\sigma^2}{|h_k|^2}\frac{\ln(2)}{B}$ in the above formula also
%leads to $t^*_k = \tau_k$ which implies $l^*_k = 0$. Also, if
%$\lambda^* > KC_kP_k + \alpha\frac{\sigma^2}{|h_k|^2}\frac{\ln(2)}{B}$,
%then the argument of $W_0(\cdot)$ is greater than $-\frac{1}{e}$
%as required.
The optimization problem can then be solved using a one-dimensional
search for $\lambda^*$, as described in Algorithm~\ref{alg:bin-search}.

\begin{algorithm}
    \algsetup{linenosize=\tiny}
    \small{%
	\caption{Binary search for $\lambda^*$}
	\label{alg:bin-search}
 	\begin{algorithmic}[1]
    \STATE $(\lambda_l, \lambda_h) = (0, \max_k\{C_kP_k +
    \alpha \frac{\Gamma \sigma^2}{|h_k^2|}\frac{\ln(2)}{B}\})$
    \STATE $(L_l, L_h) = (
    \sum_k \frac{F_k}{C_k}(\tau_k - t^*_{k,l}),
    \sum_k \frac{F_k}{C_k}(\tau_k - t^*_{k,h}))$
    where $t^*_{k,l}$ and $t^*_{k,h}$ are obtained using~\eqref{eq:opt}
    with $\lambda_l$ and $\lambda_h$, respectively.
	\WHILE {$L_l \neq L$ and $L_h \neq L$}
    \STATE $L_m = 
    \sum_k \frac{F_k}{C_k}(\tau_k - t^*_{k,m})$ where
    $t^*_{k,m}$ is obtained using~\eqref{eq:opt} with
    \mbox{$\lambda_m = (\lambda_l + \lambda_h)/2$}.
    \STATE \textbf{if} $L_m > L$ \textbf{then},
  	$\lambda_h = \lambda_m$, compute $L_h$ as in step 2.
    \STATE \textbf{else if} $L_m < L$ \textbf{then},
    $\lambda_l = \lambda_m$, compute $L_l$ as in step 2.
    \STATE \textbf{else} $\lambda^* = \lambda_m$.
  	\ENDWHILE
 	\end{algorithmic}}
\end{algorithm}

\section{Numerical results}
\label{sec:num-exp}
In this section, the performances of the optimal collaborative-computing
scheme are benchmarked against a blind collaborative-computing scheme and
the non-collaborative scenario through numerical experiments%
\footnote{Source code available at
\url{https://github.com/anpar/EE-WCC-MapReduce}.
Clicking on a figure will directly lead you to the code that generated it.}.
The blind collaborative-computing scheme simply consists in uniformly
distributing $w$ between the $K$ nodes, i.e., $l_k = L/K$, without taking
into account their computing capabilities and the strength of their channel
to the AP\@. 
Unless stated otherwise, the parameters used in the following numerical
experiments are given in Table~\ref{tab:sim-param}~\cite{EERAMECO}.

\begin{table}[h]
    \caption{Parameters used in the numerical experiments.}
    \label{tab:sim-param}
    \centering
    \begin{tabular}{clc}
        \specialrule{.1em}{0em}{0em}
        \textbf{Parameter} & \textbf{Value} & \textbf{Units} \\
        \specialrule{.1em}{0em}{0em}
        $C_k$ &
        $\overset{\text{i.i.d.}}{\sim}$ $\text{Unif}([500, 1500])$
        & [CPU cycles/bit] \\
        \hline
        $P_k$ & 
        $\overset{\text{i.i.d.}}{\sim}$ $\text{Unif}([10, 200])$
        & [pJ/CPU cycle] \\
        \hline
        $F_k$ & 
        $\overset{\text{i.i.d.}}{\sim}$ $\text{Unif}(\{
        0.1, 0.2, \dots, 1.0\})$
        & [GHz] \\
        \specialrule{.1em}{0em}{0em}
        $h_k$ &
        $\overset{\text{i.i.d.}}{\sim}$ $\mathcal{CN}(0, 10^{-3})$
        (Rayleigh fading)
        & / \\
        \hline
        $B$ & 15
        & [kHz] \\
        \hline
        $\sigma^2$ & 1
        & [nW] \\
        \hline
        $\Gamma$ & 1 
        & / \\
        \hline
    \end{tabular}
\end{table}

\subsection{Maximum computation load and outage probability}
\begin{figure}
    \centering
    % FIXME: put this back at 5.5cm if needed
    \href{https://github.com/anpar/EE-WCC-MapReduce/blob/master/src/figure2.py}{%
    \includegraphics[height=5.2cm, width=0.48\textwidth]{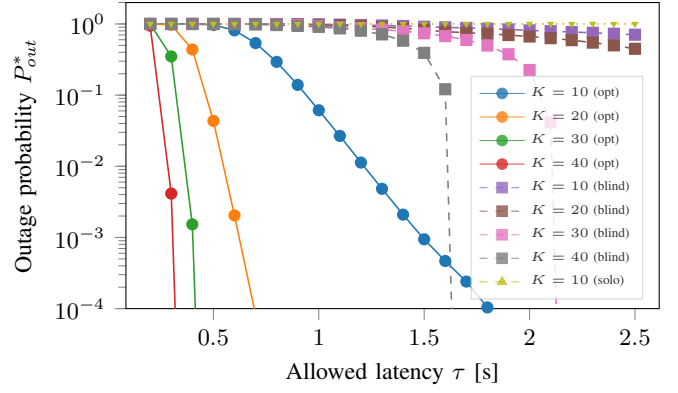}}
    \caption{%
        Empirical outage probability for the optimal
        collaborative-computing scheme (opt), the blind
        collaborative-computing scheme (blind) and the
        non-collaborative scenario (solo).
        Each data point is the result of an average on 1M experiments
        with $L =$ 4Mb, $D =$ 100b and $T =$ 5kb.
    }
    \label{fig:outage-prob}
\end{figure}

We start by looking at the maximum computation load (i.e., the maximum
size of $w$) that can be processed by the different schemes under a
given latency constraint. 
For both the optimal and the blind collaborative-computing schemes,
the maximum computation load is achieved when $\tau_k$, the effective
latency, is entirely used to perform local computation, that is, when an
infinite amount of energy is used for the Shuffling phase and
$t^\text{SHU}_k \to 0$. The maximum computation load of the optimal
and blind collaborative-computing schemes are thus given by
\begin{equation*}
    \textstyle{%
    L_\text{max}^\text{opt} = 
    \sum_{k=1}^K \frac{F_k}{C_k}\tau_k}
    \label{eq:Lmax-opt}
\end{equation*}
and
\begin{equation*}
    \textstyle{L_\text{max}^\text{blind} = K
    \min_k\left\{\frac{F_k}{C_k}\tau_k\right\},}
    \label{eq:Lmax-blind}
\end{equation*}
respectively.
For the case where the nodes do not collaborate (i.e., each node is
working for itself only), the maximum computation load that can
be processed in the allowed latency $\tau$ is given by
\[
    \textstyle{L_\text{max}^\text{solo} = \min_k\left\{\frac{F_k}{C_k}
    \left(\tau - D\frac{C_k}{F_k}\right)\right\}.}
\]
If we consider the computing capabilities of the nodes as random
variables, $L^*_\text{max}$ can also be considered as a random variable.
Thus, for a given computation load $L$, one can define the outage probability
$P_\text{out}^*$ of the system as follows
\[ P_\text{out}^* = \mathbb{P}[L^*_\text{max} < L]. \]
Figure~\ref{fig:outage-prob} shows the empirical outage probability
of the different schemes as a function of the allowed latency $\tau$
for several numbers of nodes $K$.
This figure illustrates one of the advantage of the optimal
scheme: for a given number of nodes $K$ and a given allowed latency $\tau$,
this is the scheme with the highest probability of satisfying the latency
constraint. Increasing the number of nodes is also
more profitable with the optimal scheme than it is with the blind
scheme. This is because the optimal scheme leverages \textit{diversity}
amongst the nodes, while the blind scheme, as suggested by its name,
is blind to that diversity and considers all the nodes as being equals.

\subsection{Energy consumption and energy-latency trade-off}
\begin{figure}
    \centering
    \subfloat[%
        Comparison of the total energy consumed by the optimal
        and the blind scheme for $\tau = \SI{1}{\second}$.
    ]{%
        \href{https://github.com/anpar/EE-WCC-MapReduce/blob/master/src/figures3ab.py}{%
        \includegraphics[height=5cm, width=0.48\textwidth]{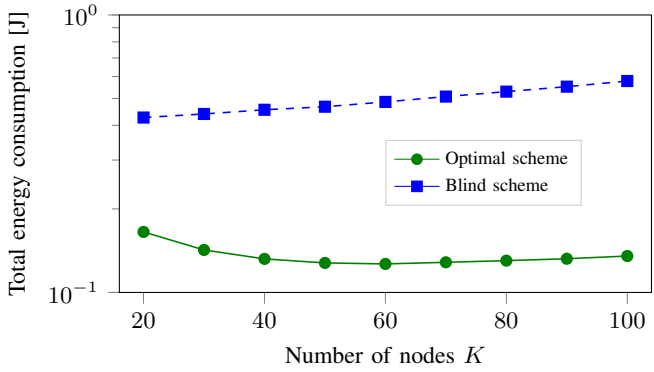}}%
    \label{fig:tot-energy}}
    \hfill
    \subfloat[%
        Breakdown of the energy consumed by the three phases of
        the collaboration, both for the optimal (opt) and the
        blind (blind) collaborative computing scheme, for
        $\tau = \SI{1}{\second}$. Note that the energy
        consumption for the Reduce phase is the same in both
        case.
    ]{%
        \href{https://github.com/anpar/EE-WCC-MapReduce/blob/master/src/figures3ab.py}{% 
        \includegraphics[height=5cm, width=0.48\textwidth]{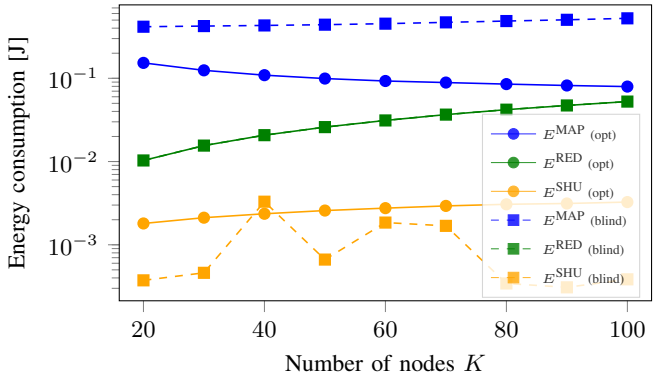}}%
    \label{fig:energy-breakdown}}
    \hfill
    \subfloat[%
        Evolution of the energy consumed by the three phases
        of the optimal collaborative-computing scheme,
        as a function of the latency, for $K = 60$.
    ]{%
        \href{https://github.com/anpar/EE-WCC-MapReduce/blob/master/src/figure3c.py}{% 
        \includegraphics[height=5cm, width=0.48\textwidth]{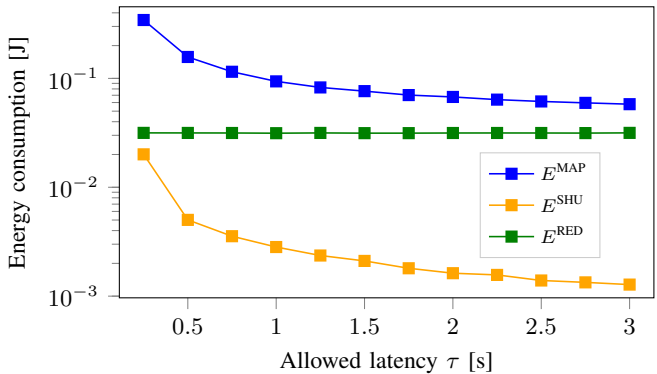}}%
    \label{fig:energy-vs-lat}}
    \caption{Energy consumption of the nodes when $L =$ 4Mb,
    $D =$ 100b and $T =$ 5kb. Each point is the result of an
    average over 10000 feasible (for each scheme) instances
    of the problem.}
    \label{fig:energy}
\end{figure}

Figure~\ref{fig:tot-energy} compares the total energy consumption
of the nodes when using the optimal and the blind scheme.
Each point on the figure is the result of an average over 10000 random
feasible (for both the optimal and the blind schemes) instances of the problem,
with $L =$ 4Mb, $D =$ 100b, \mbox{$T =$ 5kb} and with the allowed latency
$\tau$ set to $\SI{1}{\second}$ to ensure feasibility by both schemes with
relatively high probability.
This figure shows that the optimal scheme consumes approximately four to
five times less energy than the blind scheme. Note that the total energy
consumed in the non-collaborative scenario can easily be shown to be roughly
$K$ times larger than the total energy consumed by the blind scheme.
Next, Fig.~\ref{fig:energy-breakdown} breaks down the total energy
consumptions of both the optimal and the blind schemes into three
components associated to the different phases of the collaboration,
i.e., $E^\text{MAP}$, $E^\text{SHU}$ and $E^\text{RED}$.
First of all, this figure shows that most of the energy is consumed by
the Map and Reduce phases\footnote{Note that this result depends a lot
on the value of the parameters presented in Table~\ref{tab:sim-param}
and should thus not be interpreted
as a general result.}. Next, and at the opposite of the blind scheme,
the optimal scheme is able to reduce $E^\text{MAP}$ when $K$ increases,
again by leveraging diversity amongst the nodes. This explains the
slow decrease of the total energy consumption with $K$ visible on
Fig.~\ref{fig:tot-energy}. At some point however, the unavoidable
energy consumption of the Reduce phase starts to grow faster than
$E^\text{MAP}$ decreases and the total energy consumption rises again.
Finally, Fig.~\ref{fig:energy-vs-lat} depicts how the different
energy components of the optimal scheme evolve with the allowed latency
$\tau$.
In particular, this figure shows that the optimal scheme is able to
decrease the energy consumed by the Map phase when $\tau$ increases.
This is again a benefit of the diversity amongst the nodes: increasing
the allowed latency allows the optimal scheme to use slower but more
energy-efficient nodes, hence decreasing the energy consumption.

\section{Discussion and future works}
\label{sec:ccl}
In this work, an energy-efficient wireless collaborative-computing
scheme inspired by the Map-Reduce framework has been proposed.
Numerical experiments highlighted the benefits of this scheme over
a blind scheme and the non-collaborative scenario:
lower achievable latency, reduced energy consumption and the ability
to trade energy for latency and vice versa. Those benefits
are obtained by leveraging the \textit{diversity} of the nodes in
term of computing capabilities and channel strength. Analytical
results highlighting the benefits of diversity are however missing
and their pursuit thus constitutes a first possible direction for
future works.

A second obvious direction for future works might be to refine the
optimization problem formulated in Sec.~\ref{sec:problem-formulation}
to account for additional constraints, e.g., limited memory capacity,
maximum RF transmit power or limited battery level. 

Next, the models used in this work to quantify the time and energy
consumed by the different phases of the collaboration are
very simple and far from being realistic (see, for
instance,~\cite{bouguera2018energy}). Incorporating more realistic
models in the proposed collaborative-computing scheme will thus
certainly be a priority in future works.

Finally, as opposed to the original Map-Reduce framework
that considers some redundancy between the smaller files $\{w_k\}_{k=1}^K$
to increase the robustness of the system to node failure and to prior
works~\cite{scalableWDC, coding-edge, comm-comp-tradeoff, wireless-mr}
that study the trade-off between computation and communication load
through network coding, we did not assume any redundancy in this work.
Investigating the possible benefits of redundancy in the proposed scheme
is thus another interesting research question.

\section*{Acknowledgment}
AP is a Research Fellow of the \textsc{F.R.S.-FNRS}\@.
This work was also supported by \textsc{F.R.S.-FNRS}
under the \textsc{EOS} program (project 30452698,
``MUlti-SErvice WIreless NETwork'').

\bibliographystyle{IEEEtran}
\bibliography{main.bib}

\end{document}